\documentclass[runningheads]{llncs}
\usepackage{graphicx}
\usepackage{cite}
\usepackage{color}
\usepackage{caption}
\usepackage{placeins} 
\usepackage{multirow} 
\usepackage{multicol} 
\usepackage{amsmath}
\usepackage{amssymb} 
\usepackage[ampersand]{easylist} %
\usepackage{xspace}
\usepackage[inline]{enumitem}
\usepackage{tabularx}
\ListProperties(Hide=100, Hang=true, Progressive=3ex, Style*=-- ,Style2*=$\bullet$ ,Style3*=$\circ$ ,Style4*=\tiny$\blacksquare$ ) 
\usepackage[inline]{enumitem}
\usepackage{subfig}
\usepackage{todonotes} 

\newcounter{bean}

\newcommand{\Omit}[1]{}

\makeatletter
\newcommand*{\rom}[1]{\expandafter\@slowromancap\romannumeral #1@}
\makeatother 
\newcommand\kw[1]{\texttt{#1}\xspace}
\newcommand{\male}{\kw{Male}}
\newcommand{\female}{\kw{Female}}
\newcommand{\married}{\kw{Married}}
\newcommand{\loneparent}{\kw{Lone parent}}%
\newcommand{\youngchild}{\kw{Dependent under 15 child}}
\newcommand{\student}{\kw{Dependent student}} 
\newcommand{\adultchild}{\kw{Non-dependent over 15 child}} 
\newcommand{\children}{\kw{Children}}%
\newcommand{\relative}{\kw{Relative}}%
\newcommand{\grouphh}{\kw{Group household}}%
\newcommand{\loneperson}{\kw{Lone person}}%

\newcommand{\hh}{\kw{Household}}%
\newcommand{\family}{\kw{Family}}%
\newcommand{\familytypes}{\kw{Couple family with no children}, \kw{Couple family with children}, \kw{One parent family} and \kw{Other family}}

\begin{document}
%
\title{Building a large synthetic population from Australian census data}
\titlerunning{Synthetic population}
%

\author{Bhagya N. Wickramasinghe\inst{1} \and  Dhirendra Singh\inst{1} 
\and
 Lin Padgham\inst{1}}
\authorrunning{B. N.Wickramasinghe et al.}
\institute{RMIT University, Melbourne, Australia\\
\email{first.last@rmit.edu.au}\\
}

\maketitle              

\begin{abstract}
We present work on creating a synthetic population from census data for Australia, applied to the greater Melbourne region.
We use a sample-free approach to population synthesis that does not rely on a disaggregate sample from the original population.
The inputs for our algorithm are joint marginal distributions from census of desired person-level and household-level attributes, and outputs are a set of comma-separated-value (.csv) files containing the full synthetic population of unique individuals in households; with age, gender, relationship status, household type, and size, matched to census data.
Our algorithm is efficient in that it can create the synthetic population for Melbourne comprising 4.5 million persons in 1.8 million households within three minutes on a modern computer.
Code for the algorithm is hosted on GitHub.

\end{abstract}

\section{Introduction} \label{sec:intro}

A synthetic population is an algorithmically generated virtual population that matches some known views, such as those given by census tables, of a real population. The aim of population synthesis is to construct instances of persons and households and assign to them attributes of the real population, such as age, sex, and relationships, in such a way that the virtual population is a good match, statistically, to the real population, with respect to those attributes.
Synthetic populations are often used in agent-based simulations so real-life phenomena can be replicated and ``what-if'' scenarios explored. Our own effort is driven by ongoing work with emergency services in modelling community evacuations to improve preparedness for bushfires and floods in Australia. This work also fits in a larger project with social science researchers looking at developing an understanding of how households make housing decisions, to answer questions around what kind of housing stock we should build to get the kinds of communities we aspire to in Plan Melbourne 2050\footnote{http://www.planmelbourne.vic.gov.au/}.

We use a sample-free approach to population synthesis that does not rely on a disaggregate sample from the original population to \textit{seed} the algorithm, as is required by established methods like Iterative Proportional Updating (IPU)~\cite{Ye2009}. This is particularly useful in the Australian context where samples are not freely available and come with substantial restrictions on use. Joint marginal distributions for exactly the desired attributes of the population can readily be downloaded from the Australian Bureau of Statistics (ABS) website\footnote{\url{https://abs.gov.au}}, making it unnecessary to use techniques like Iterative Proportional Fitting (IPF)~\cite{ipfp1,ipfp2} for deriving such. Our work is more closely related to that of Huynh et. al~\cite{Huynh2016} that also employs a sample-free approach for Australia, but uses a different mechanism for assigning individuals to households. We also use the more recent census data from 2016 (catering to any changes in data format) and report better accuracy on the match to census data.

Our approach can be used to build a synthetic population for any area of
Australia covered by the census.
The program outputs the population at the granularity of Statistical
Area Level 2 (SA2) that represents a community with about 10,000
residents, and typically corresponds to officially gazetted state suburbs and
localities.\footnote{\scriptsize\url{www.abs.gov.au/websitedbs/D3310114.nsf/
home/Australian+Statistical+Geography+Standard+(ASGS)}}
We demonstrate the approach using 2016 census
data\footnote{We have also tested on ABS 2011 census data.} for all 309 SA2s of
Greater Melbourne consisting of 4,485,211 persons in 1,832,043
households\footnote{\scriptsize\url{www.censusdata.abs.gov.au/census%
\_services/getproduct/census/2016/quickstat/2GMEL}}.
The program takes about 3 minutes to perform the data cleaning
routines and generate the population
on a 64-bit computer with 8 $\times$ 3.40GHz
cores and 8 GB RAM (Intel\textsuperscript{\textregistered}
Core\textsuperscript{\texttrademark} i7-4770 CPU).
The full source code for the algorithm, all supporting scripts, instructions for running the
program, as well as the final generated population files for Melbourne are
available from the GitHub 
repository\footnote{\url{www.github.com/agentsoz/synthetic-population}}.

The remainder of the paper is laid out as follows. Similarities and differences to other related works are first covered in Section~\ref{sec:related}. We then describe our algorithm in Section~\ref{sec:approach} and compare the quality of the produced synthetic population to the original census data in Section~\ref{sec:eval}. We further compare our approach in detail to relevant existing techniques in Section~\ref{sec:comparison} and conclude with a discussion on nuances in Section~\ref{sec:conclusion}.

\section{Related Work} \label{sec:related}


There is substantial work on population synthesis for agent based
models~\cite{Hermes2012}. Much of this focuses on creating individuals within
households based on census data, as we do here. In many countries the required
joint distributions are not directly available from census data, and techniques
like Iterative Proportional Fitting (IPF)~\cite{ipfp1,ipfp2} are used to
construct the desired joint aggregated (marginal) distributions using a set of
available marginals and a disaggregate data \textit{sample} (microdata). 
Other sampling techniques that have been proposed include the Gibbs sampler with
Markov Chain Monte Carlo simulation~\cite{Farooq2013} and Quasi-random integer
sampling~\cite{Smith2017}.

Of the sample-based approaches,
IPF is a widely used deterministic re-weighting technique to merge data
distributions at the same aggregation level.
It takes a set of marginal distributions
and a disaggregate data sample from the underlying population (the
seed) and iteratively adjusts (reweights) the seed to match the
marginal distributions~\cite{Beckman1996}.  GREGWT (Generalised Regression and
Weighting)
is also an iterative re-weighting algorithm used by the Australian Bureau of
Statistics~\cite{Tanton2011}. Iterative Proportional Update
(IPU)~\cite{Ye2009} and Hierarchical IPF~\cite{Kirill2011} are two
deterministic re-weighting techniques similar to IPF but designed
to merge data distributions of different aggregation levels,
e.g., person and household level distributions. The sample
data used in these works represent the population at both aggregation
levels.

Combinatorial optimisation techniques~\cite{Hermes2012} use two marginal 
distributions at person and household levels and a sample from the underlying
population containing household structures as inputs. The process
starts by generating an initial estimate of the whole population by
cloning households with persons from the data sample. This approach
ensures realistic household structures and accurate relationships
within the household. The goodness of fit of the estimate is then
optimised against an objective function using a combinatorial
optimisation method like hill climbing, simulated annealing or genetic
algorithms~\cite{Williamson1998,Ballas2003,NamaziRad2014a}.
The main drawback of these approaches is possible loss
of heterogeneity due to repeated cloning from a small subset of
households~\cite{Farooq2013}.

The above methods assume availability of a disaggregate data
sample, which is not always available or if available may be
undesirable due to restrictions on use.  Alternative
\textit{sample-free} methods
instead employ heuristics to infer household structures~\cite{Huynh2016}.
Additionally, \cite{Ye2017} show that populations of the
same aggregation level (person or household level) can be
constructed without a sample when joint marginal distributions with
sufficiently overlapping attributes are available.

Yet another approach is to generate
household templates by taking all combinations of person and
household types, and using them to perform Monte Carlo sampling
considering known person and household level
distributions.  However, the number
of combinations grows exponentially in the number of person and
household types, as shown by~\cite{Gargiulo2010} for a region in France.
We evaluated this approach also for Melbourne with
2 household attributes with 8 and 14 values respectively giving 112 categories,
and 3 person attributes with 2, 8 and 8 values, giving 172
categories. On a 14 core (2.6 GHz) supercomputer with 1TB RAM however,
the process failed to complete after 48 hours.

Due to the computational complexity of the above approach
heuristics are typically used to place individuals into
households.
In this approach household instances are formed by
selecting a household and suitable persons from corresponding pools,
without replacement, according to population
heuristics~\cite{Barthelemy2013,Huynh2016}.
This is the general approach we have used.
We compare our approach to that of~\cite{Huynh2016} as well as
to the well-known IPU technique~\cite{Ye2009} further in
Section~\ref{sec:comparison}.
\section{Population Synthesis Approach} \label{sec:approach}

The complete population synthesis process involves downloading the
input person-level and household-level marginal distributions from the
ABS website using the
TableBuilder\footnote{\scriptsize\url{www.abs.gov.au/websitedbs/censushome.nsf/home/tablebuilder}}
software tool;
running the pre-processing script that performs some data cleaning
steps and generates the required counts for the different person and
household categories according to the downloaded marginal distributions; and
running the population synthesis algorithm (Section~\ref{subsec:latch-algo}).

The output of the program is the population of each SA2 in three
comma-separated-values (.csv) files containing person, families (of persons), and households (of families). Each record (row) in the
household file primarily contains a unique ID for the household, the
household size, the family household composition of the primary family
and an ID representing the primary family. A record in the families file
represents each family with a unique ID, the family household
composition category of the family and the household ID it belongs
to. A record in the persons file represents persons with unique IDs, their
age, sex, relationship status, family ID, household ID and the IDs of
their partner, father, mother, children and relatives. Together the columns in the output files
constitute a valid database schema that could easily be loaded and manipulated in any database management system.


\subsection{Person-level Categories Input}\label{subsec:census-ind-data} %

Person-level attributes that we chose to include in the synthetic population
consist of eight age categories (\kw{0-14}, \kw{15-24}, \kw{25-39}, \kw{40-54},
\kw{55-69}, \kw{70-84}, \kw{85-99}, and \kw{100 and over}), two gender
categories (\male and \female) and eight family/household relationship
categories that we obtained by aggregating the census categories
as per Table~\ref{tab:census-indiv-orinal-custom-map}. This gave us
128 customised person categories (2$\times$8$\times$8).
Note that not every custom person category is valid, for instance an individual
in the \kw{0-14} age bracket cannot have the \married relationship status. The
total number of legal person-level categories are actually
90.

\married is assumed to imply a heterosexual relationship for simplicty, even though census data includes also homosexual marital partnerships.
Children are categorised
as \youngchild (aged 0-14), \student (aged 15-24, studying full-time, living 
with
parents), or \adultchild (aged 15 or over, not studying full-time, living with
parents).
The age bins above age 24, e.g., \kw{25-39} are arbitrarily chosen.
~\relative is a related individual who lives in a family household but is not part of the
family nucleus, or forms a family nucleus from a non-marital/non-parental relationship, e.g., siblings living in the same household. \grouphh includes
individuals living together with other non-related individuals, e.g.,
tenants in a shared house, while a \loneperson lives alone.

\begin{table}[h]%
	\caption{Custom relationship status categories based on classifications
	used by
		Australian Bureau of Statistics}%
	\centering %
	\resizebox{\textwidth}{!}{ \begin{tabular}{|l|l|} %
			\hline %
			\textbf{Custom category} & \textbf{Original categories}\\%
			\hline %
			\married & Husband, Wife or Partner in de facto marriage, female
			same-sex
			couple\\%
			&Husband, Wife or Partner in de facto marriage, male same-sex
			couple\\%
			&Husband, Wife or Partner in de facto marriage, opposite-sex
			couple\\%
			&Husband, Wife or Partner in a registered marriage\\%
			\hline%
			\loneparent & Lone parent\\%
			\hline %
			\youngchild&Foster child under 15\\%
			&Grandchild under15\\%
			&Natural or adopted child under 15\\%
			&Otherwise related child under 15\\%
			&Step child under 15\\%
			&Unrelated child under 15\\%
			\hline%
			\student& Natural or adopted dependent student\\%
			&Dependent student step child\\%
			&Dependent student foster child\\%
			\hline%
			\adultchild & Non-dependent foster child\\%
			&Non-dependent step child\\%
			&Non-dependent natural, or adopted child\\%
			\hline%
			\relative & Brother/Sister\\%
			&Cousin\\%
			&Father/mother\\%
			&Grandfather/grandmother\\%
			&Nephew/niece\\%
			&Non-dependent grandchild\\%
			&Other related individual (nec)\\%
			&Uncle/aunt\\%
			&Unrelated individual living in family household\\%
			\hline%
			\grouphh & Group household member\\%
			\hline%
			\loneperson & Lone person \\%
			\hline%
			\textit{Ignored}& Visitor(from within Australia)\\%
			&Not applicable\\%
			&Other non-classifiable relationship\\%
			\hline %
		\end{tabular}	%
	} %
	\label{tab:census-indiv-orinal-custom-map} %
\end{table}%

In the input data, where a person could be assigned one of several relationship types, the ABS uses certain rules to determine the relationship status, such as in order of priority: marital, then parental/child relationship, then relative. In case of multi-generation
parent/child relationships, the younger one is given the higher priority after
any marital relationships. Any adult not belonging to a family nucleus is
categorised as a relative.

\subsection{Household-level Categories Input}\label{subsec:census-hh-data}

Structurally in the ABS data, a \hh is composed of either 1$\times$\loneperson,
1$\times$\grouphh, or 1-3$\times$\family units where a \family consists of at
least two
persons and each person belongs to one \family unit alone. The \hh also
specifies the type of the \textit{primary} family unit, which is one of:
\familytypes; while no information is provided on the non-primary family types.
Finally, the \hh structure has a \text{size} that gives the total number of
persons in the complete household. Together, this gave us 112 household
categories: 8 size categories (\texttt{[1,2,\ldots,8+]}) $\times$ 14 household
unit types (\loneperson $+$ \grouphh $+$ [1/2/3-\family $\times$ 4 primary
family types]). As with person-level categories, not all of the household-level
categories are valid, for instance, a household with two persons cannot be a
3-\family household, and the total number of valid categories here is
65.

The ABS uses rules to assign family types in the data when one of several
below types apply, such as: marital only as \kw{Couple only}, marital and parent/child as
\kw{Couple family with children}, parent/child only as \kw{One parent family} and relative
relationships only as \kw{Other family}. Except for \kw{Other family}, relative relationships are ignored when deciding the family type.
In multi-family households, the primary family
is selected in the prioritised order of \kw{Couple family with children}, then \kw{One parent
family} and finally \kw{Couple only} or \kw{Other family} with equal priority.


\subsection{Population Synthesis Algorithm}\label{subsec:latch-algo}

We  construct the population in five stages. %
The first is to instantiate all the persons instances with their
attributes.
Next is to form the basic structures for all the inferable families with the minimum
persons required as per the family type. We can infer that all the
\loneparent{}s must form basic \kw{one parent} families with a child
because \loneparent{}s cannot be in any other family type, and all the \married
\male{}s and \female{}s must form couples. Further based on the households data
distribution we can form basic structures needed for primary \kw{Couple
  Family with Children} families by pairing a previously formed couple with a
child and primary \kw{Other Family} units by pairing two
\relative{}s.
The third stage is instantiating all the family households with the correct
primary family using the basic family instances, \kw{Group Households}
with \grouphh persons and \kw{Lone Person} households with \loneperson
instances.
The fourth stage is heuristically adding non-primary families to the incomplete
family households. After this step all the family households have required
families and some may even have all the required members.
Finally, the remaining incomplete households are completed by adding remaining
\children and \relative{}s.
As an optional last step, we can assign households in a SA2 to known street addresses in that SA2 (assigned arbitrarily) which can be useful for geography-based applications.
We describe the full algorithm in steps below.


\newcounter{stepno}
\newcommand{\sitem}{\refstepcounter{stepno}{\noindent\bf S\thestepno}---}


  \subsection*{Stage 1 | Create All Persons}\label{subsec:create-persons}
  \sitem Clean the input distributions obtained from TableBuilder.
  Discrepancies exist between the household and
  individual data obtained either due to limitations in
  the data collection process or random errors deliberately introduced by ABS in the census data to
  protect privacy\footnote{\scriptsize\url{http://www.abs.gov.au/ausstats/abs@.nsf/Lookup/2901.0Chapter38202016}}.
  We make adjustments to remove these discrepancies to
  the extent possible
  including ensuring the number of
  individuals in each set are the same, ensuring there are enough
  married males and females to create the couple families and so on. We
  treat the household-level data as the most correct, and adjust the
  individual-level data to match this.

  \sitem Create the required number of person instances and assign their attributes as per the input person-level categories distribution.
  At this stage a person is assigned only an age bin. Once the algorithm has run to completion,
  we convert these to an age by stochastically selecting an age from the age distributions of persons in the SA2. This also
  considers \textit{population heuristics} when selecting potential age values
  for a person to avoid any unrealistic relationships between
  persons. For example we require the parent-child age gap
  to be at least 15 years but not more than 45 years.
  \label{latch:algostep:assign-absolute-age}%
  The formed instances are added to separate pools as:
  \textit{married-males}, \textit{married-females}, \textit{lone parents},
  \textit{children}, \textit{relatives}, \textit{group-households} and
  \textit{lone persons}.

  \sitem If there are more persons represented in household-level
  compared to the individual-level distribution, they are instantiated without
  any attributes and added to the \textit{extras} pool. These extras are
  used later in population construction if any person type does not have enough
  instances to form the required households. %

  \subsection*{Stage 2 | Form All Known Basic Family Structures}
  \sitem Form basic structures of \kw{One parent} families by pairing each
  person in the \textit{lone parents} pool with a person from the
  \textit{children} pool in age descending order, adhering to the parent-child age gap
  constraint. The constructed
  family structures are put in the \textit{basic one parent} family units pool.
  \label{latch:algostep:forming-basic-one-parent}%

  \sitem Form all the possible married \textit{basic couples} by
  pairing two persons, each taken from the
  \textit{married-males} and the \textit{married-females} in the age descending
  order, until one or both pools is depleted. Any remaining persons will be used in later
  stages. All marital relationships are
  assumed to be	heterosexual.\label{latch:algostep:forming-basic-couple-family} %

  \sitem Form the pool of \textit{basic couple with child} family units by
  grouping a randomly selected unit from the \textit{couples} pool and a child from
  the \textit{children} pool, to match the number of households in the input
  household distribution where the primary family is of type \kw{Couple
    Family with Child}. The parent-child age gap constraint applies here as before.
  \label{latch:algostep:forming-basic-couple-w-child}%

  \sitem Form \textit{basic other family} units by grouping two randomly selected
  \relative persons to match the number of households where the primary family is of
  type \kw{Other Family} according to the input household
  distribution.\label{latch:algostep:forming-basic-other-family} %

  \subsection*{Stage 3 | Create All Households} %

 \sitem Create all households and assign the desired attributes:
 household size, primary family type, and number
 of family units, as per the household-level distribution and add them to the
 \textit{incomplete-households} pool.%

 \sitem Add each individual in the \textit{lone-persons} pool to each
 \kw{Lone person} households in the input households distribution and add
 the households to the \textit{completed-households}. %

 \sitem Complete the \kw{Group households} by adding all the required person
 instances from the \textit{group-households} pool and add the constructed households
 to the \textit{completed-households} pool.%

 \sitem Assign the primary family to each household in
 \textit{incomplete-households} by selecting a family unit that matches the
 household's primary family type from one of the basic family unit pools created
 in steps~\ref{latch:algostep:forming-basic-one-parent},
 \ref{latch:algostep:forming-basic-couple-family},
 \ref{latch:algostep:forming-basic-couple-w-child} and
 \ref{latch:algostep:forming-basic-other-family}. This step utilises all the
 units in the \textit{basic other family} and the \textit{basic couple with
   child} pools, as these pools only have families sufficient for primary
  families. %

  \subsection*{Stage 4 | Assign Non-primary Families}%
  The following criteria applies to the selection of non-primary families: %
  \begin{enumerate*}[label=(\alph*)]
    \item to add any family type there must be at least one free slot for a new
    non-primary family\label{latch:cond:nonprime-morefamily};
    \item to add a \texttt{Couple family with no children} or a \texttt{Other
      family} units there must be room for at least two more
    persons\label{latch:cond:nonprime-coupleother}; %
    \item to add a \texttt{One Parent} family there must be room for at least two
    persons and the primary family must be either \texttt{Couple Family with
      Children} or \texttt{One Parent}
      family\label{latch:cond:nonprime-oneparent}; and%
    \item to add a \texttt{Couple Family with Children} unit there must be room
    for at least three persons and the primary family type must also be
    \texttt{Couple Family with Children}\label{latch:cond:nonprime-cplwchild}. %
  \end{enumerate*}\\

 \sitem Assign remaining units in the \textit{basic one parent} pool to randomly
 selected eligible households in \textit{incomplete-households} pool that meet
 criteria \ref{latch:cond:nonprime-morefamily} and
 \ref{latch:cond:nonprime-oneparent} above until depleted or no eligible
 households remain. The same household may be selected more than once as long as
 it meets the eligibility criteria. Any completed households as a result are
 added to \textit{completed-households}.

 \sitem Disassemble any remaining units in the \textit{basic one parent} pool
 and add the persons to the \textit{children} and the \textit{lone parent}
 pools accordingly.

 \sitem Add remaining units in \textit{basic couples} pool to randomly selected
 households in the \textit{incomplete-households} that meet conditions
 \ref{latch:cond:nonprime-morefamily}, \ref{latch:cond:nonprime-coupleother}
 and \ref{latch:cond:nonprime-cplwchild}. If a household only meets conditions
 \ref{latch:cond:nonprime-morefamily} and \ref{latch:cond:nonprime-coupleother}
 the new family is added as a \kw{Couple family with no children} unit. If
 the household meets conditions \ref{latch:cond:nonprime-coupleother} and
 \ref{latch:cond:nonprime-cplwchild} the new family is either assigned as a
 \kw{Couple Family with Children}, by adding a new child to it,  or a
 \kw{Couple Family with no Children} unit based on a user defined probability. 
 Fully formed households are
 added to \textit{completed-households} as before.

 \sitem Disassemble any unassigned families in \textit{basic couples} and
 re-add those persons to the \textit{married-males} and the
 \textit{married-females} pools accordingly.%

 \sitem Calculate the probability distribution $\rho$ of the primary family unit having
 \textit{marital, parental} or \textit{other} as the main relationship.

 \sitem For every household in the \textit{incomplete-households} that meets
 condition \ref{latch:cond:nonprime-morefamily}, stochastically select the
 relationship between the main two persons in the new family using $\rho$.
 \label{latch:algostep:unknown-start}
 The new family type is then given by the following table:\newline
 \begin{tabular}{p{0.27\textwidth}| p{0.16\textwidth} | p{0.5\textwidth}}
\hline
 Selected relationship & Criteria met & New family type \\\hline
 \textit{parental} & \ref{latch:cond:nonprime-oneparent} & \kw{One Parent}\\
 \textit{other} & \ref{latch:cond:nonprime-coupleother} & \kw{Other Family}\\
 \textit{marital} & \ref{latch:cond:nonprime-coupleother} & \kw{Couple with no Children}\\
 \textit{marital} & \ref{latch:cond:nonprime-cplwchild} & \kw{Couple Family 
 with Children} or \kw{Couple Family with no Children} chosen using user 
 defined probability\\\hline
 \end{tabular}

  \sitem Form a family with the basic structure for the determined family type
  according to steps~\ref{latch:algostep:forming-basic-one-parent}--\ref{latch:algostep:forming-basic-other-family}. If there are insufficient persons
  of any required relationship status, new persons are drawn from the
  \textit{extras} pool. The age and sex properties of the new persons are set
  probabilistically based on the input persons distribution considering
  population heuristics. \label{latch:algostep:nonprimary-onep}%

  \sitem Assign the new family to the household. If the household becomes
  complete add it to \textit{completed-households}. \label{latch:algostep:uknown-end}%

  \sitem Repeat steps \ref{latch:algostep:unknown-start}--\ref{latch:algostep:uknown-end} until all non-primary families are assigned.%

\subsection*{Stage 5 | Complete Households with Children and Relatives} %

\sitem Randomly select the households in the \textit{incomplete-households} where
the primary family type is \kw{One Parent} family or \kw{Couple Family
  with Children} and assign them persons from the \textit{children} pool, while
adhering to the parent-child age gap constraint. Children are only added to the
primary family to ensure it has more children than the others. Completed households
as a result are added to \textit{completed-households}.%

\sitem If persons remain in the \textit{children} pool for not having a 
suitable primary family add them to
\kw{Couple Family with Children} and \kw{One Parent} secondary and
tertiary families considering the parent-child age gap constraint and ensuring
that the updated family stays smaller than the preceding families. Completed
households are added to \textit{completed-households}.%

\sitem Convert all the remaining persons in \textit{married males},
\textit{married-females}, \textit{lone parents} and \textit{children} to the
\textit{extras} pool by nullifying their relationship status but not age and sex
categories. %

\sitem Add persons in the \textit{extras} pool to the primary family of
households in \textit{incomplete-households} until the household reaches its
expected size. The properties of the new person is determined probabilistically
based on the distribution of \youngchild, \student, \adultchild and \relative
persons in the person-level distribution and considering the population
heuristics.
If there are persons in \textit{extras} that match
the decided age and sex categories use them by setting the correct relationship
status, otherwise, all the properties are set to the expected values. Add the
completed households to \textit{completed-households}.%

\sitem Complete the population by assigning \relative{} to the remainder of
\textit{incomplete-households} until each household reaches its
expected household size. Relatives are only added to the primary family of
the household.\label{latch:algostep:end}\\~\\
Steps 1--\ref{latch:algostep:end} generate the population of persons, families, and households, for one SA2 area. The process is repeated for every SA2 in Greater Melbourne at which point the population synthesis is complete.


\subsection*{Stage 6 (Optional) | Assign households to dwellings}

The population we have constructed is accurate at the granularity of a suburb 
(SA2). To assign households to street addresses, one simple approach could be 
to arbitrarily allocate households to known street addresses in the given SA2. 
Our approach however is to do slightly better, and get the distribution of 
households correct to the more finer granularity of SA1--roughly the size of 
one block with about 400 
persons\footnote{\scriptsize\url{www.abs.gov.au/websitedbs/D3310114.nsf%
/home/Australian+Statistical+Geography+Standard+(ASGS)}}.

We do this in two stages. First, for a given household, we figure out the most suitable SA1 within the given SA2 to assign the household to. To do this we download the distribution of household counts by household type and SA1 for each SA2 from ABS TableBuilder (and perform a preprocessing step to ensure that the SA1 household counts correctly match the SA2 household counts), and using this, randomly distribute the household instances in the synthesised population among the SA1s considering their household type. Then in the second stage we assign the household to a known street address in the allocated SA1.

The process of constructing the list of street addresses for a each SA1 is the 
most expensive step in the entire process, since it requires a geometric 
calculation to determine whether a point (the coordinates of a street address) 
lies within a polygon shape (the SA1 boundary), and given that there are 
several hundred thousand street addresses in Melbourne, and several thousand 
SA1s. Note that this process need only be performed once, or whenever new 
street addresses need to be incorporated, which is not very often\footnote{The 
street addresses data we use is obtained from Vicmap (www.land.vic.gov.au) in 
ESRI Shapefile format and does not change frequently.}. Nevertheless, a naive 
combinatorial calculation is still very expensive and the computation takes 
several days on our test machine. To overcome this issue, we devise a two-pass 
process that is extremely fast. In the first pass, we do a quick calculation of 
whether an address point \textit{possibly lies} within a given SA1 polygon by 
computing whether the point lies within the bounding box for the polygon. If 
the point does not lie inside the bounding box then we can be sure that it does 
not belong to that SA1. Only if it does do we perform the more expensive 
calculation of computing whether it actually lies within the SA1 polygon. This 
tremendously reduces the computation required, since a first-pass check for 
each address against each SA1 bounding box yields 2-4 possibilities that have 
to be checked in the second pass, compared to every SA1 being checked in the 
naive method. The two-pass method computes the full map of street addresses to 
SA1s in around 24 minutes on our test machine\footnote{Specifications of the 
machine are provided in Section~\ref{sec:intro}.}.

\section{Similarity to census data} \label{sec:eval}


We evaluate the quality of the resulting synthetic population using 
cosine similarity which is a measure of similarity between two non-zero
vectors~\cite{Salton1983}. It is widely used in the information retrieval domain to
measure the similarity between documents. For example, to measure the
similarity of two documents, one would obtain a vector of unique word counts
from each document and apply cosine similarity to obtain a value between 1 (high correlation) and
0 (no correlation). In a similar way, in our context
we measure the similarity between the
expected and the observed marginal distributions, which are discrete
categorical distributions similar to the document comparison case.
Cosine similarity is calculated as:
\[
\textrm{Cosine similarity} = \frac{ \sum\limits_{i}
O_{i}E_{i}}{\sqrt{\sum\limits_{i}
O_{i}^{2}}\sqrt{\sum\limits_{i}
E_{i}^{2}}}
\]
where $O$ is the observed distribution (vector) and $E$ is the expected
distribution.

Cosine similarity measures the angle between the two vectors on a Cartesian
plane. If the two vectors coincide the angle is $0^\circ$ and the corresponding
cosine value is 1. It is also important to note that cosine does not take into
account the magnitude of the two vectors, that means if the two populations are
different in size but has the same distribution cosine similarity would report
values close to 1. This behaviour has two main implications on population
synthesis applications:
\begin{enumerate*}[label=(\itshape\alph*\upshape)]
\item when using cosine similarity the size of the expected and observed
populations need to reported to prove that there is no significant difference
in population size and
\item cosine similarity is robust at the presence of
unavoidable size differences between the expected and the observed, as long as
differences are reasonable.
\Omit{and
\item when synthesising abstract populations
which are inherently different in size, explicit normalisation
is not required.}

\end{enumerate*}

\subsubsection{Results}

For each of the 306 SA2s, we measured cosine similarity for person-level and household-level distributions. As well, we evaluate the age assignment from age bins (Section~\ref{subsec:create-persons} Step~\ref{latch:algostep:assign-absolute-age}) by also calculating the cosine similarity for the distribution of ages achieved.
Table~\ref{tab:latch:cosine-summary} shows the cosine similarity results for the three distributions we tested.
The number of SA2s that had cosine values greater than 0.99 were:
301/306 (or 98.4\%) for person-level distribution, 306/306 (or 100\%) for
household-level distribution, and 18/306 (or 5.9\%) for age  
distribution (98.4\% were however greater than 0.9 for age
distribution - still quite strong similarity). %
In all SA2s the total number of households were similar in both synthesised and 
census distributions, while total number of persons only exhibited about 2\% 
difference on average.

\begin{table}
  \centering%
  \caption{Cosine similarity (CS) results for 306 SA2s in Melbourne}%
  \label{tab:latch:cosine-summary}
  \begin{tabular}{|p{0.35\textwidth}|p{0.3\textwidth}| p{0.3\textwidth}|}
    \hline
    {\bf Distribution tested} & {\bf SA2s with CS $>$ 0.90} & {\bf SA2s with CS $>$ 0.99}\\\hline\hline
        Person (90 categories) & 303 (99\%) & 301 (98.4\%)\\
        Household (65 categories)& 306 (100\%)  & 306 (100\%) \\
        Age (101 categories) & 301 (98.4\%) & 18 (5.88\%) \\
        \hline
    \end{tabular}
\end{table}

\begin{table}[h]
  \centering%
  \caption{Census 2016 population for SA2s with cosine $<$ 0.9}
  \label{tab:latch:cosine-weak-sa2s}
  \begin{tabular}{|p{0.4\textwidth}|p{0.3\textwidth}|p{0.3\textwidth}|}
    \hline
    {\bf SA2} & {\bf Total Persons} & {\bf Total Households} \\\hline\hline
        Braeside & 12 & 9 \\
        Melbourne Airport & 60 & 17 \\
        Port Melbourne Industrial & 3 & 5 \\\hline
    \end{tabular}
\end{table}

\begin{figure}[ht]
  \subfloat[Person-level\label{subfig:yarraville_qq_person}]{
    \includegraphics[width=0.3\textwidth]{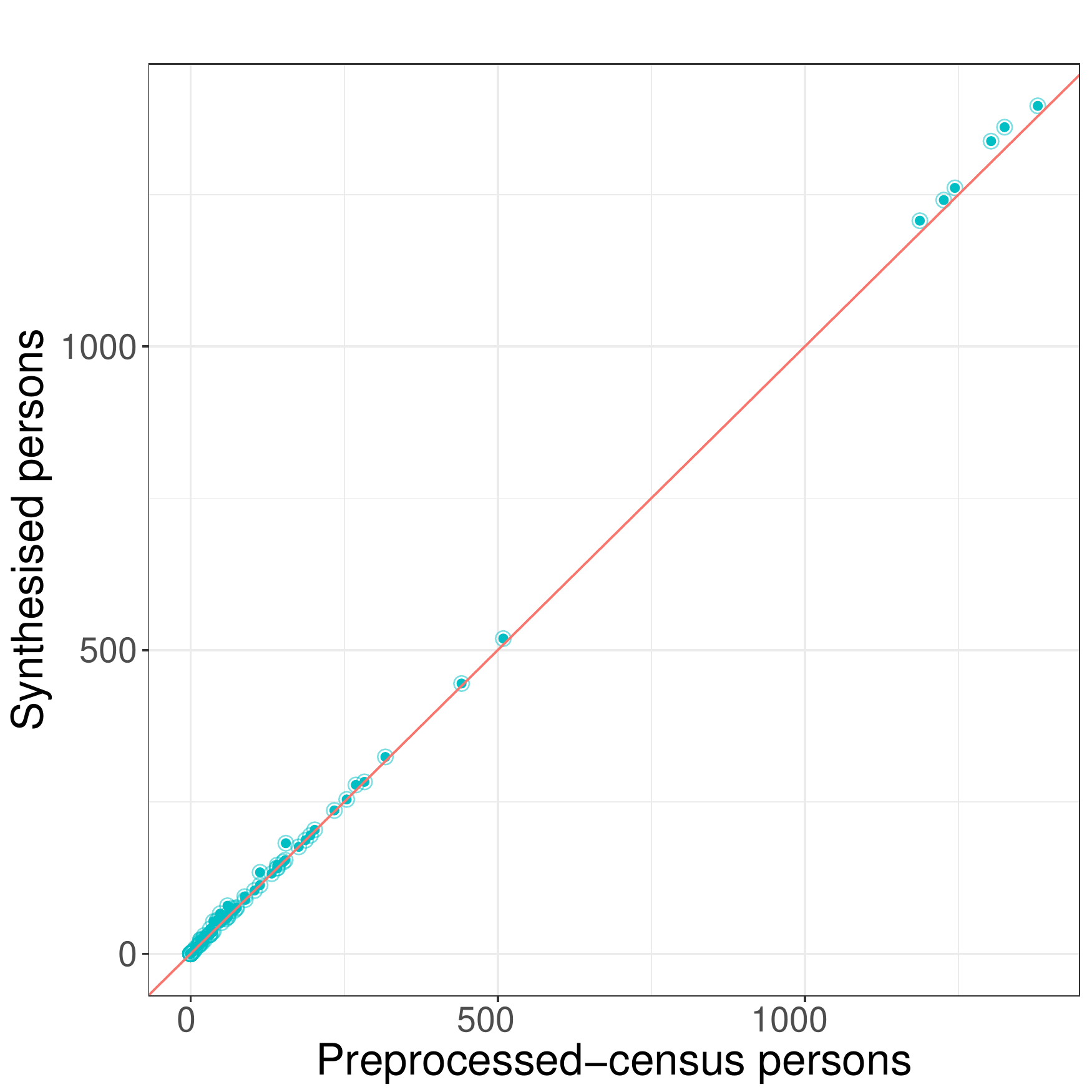}
  }
  \hfill
  \subfloat[Household-level\label{subfig:yarraville_qq_HH}]{
    \includegraphics[width=0.3\textwidth]{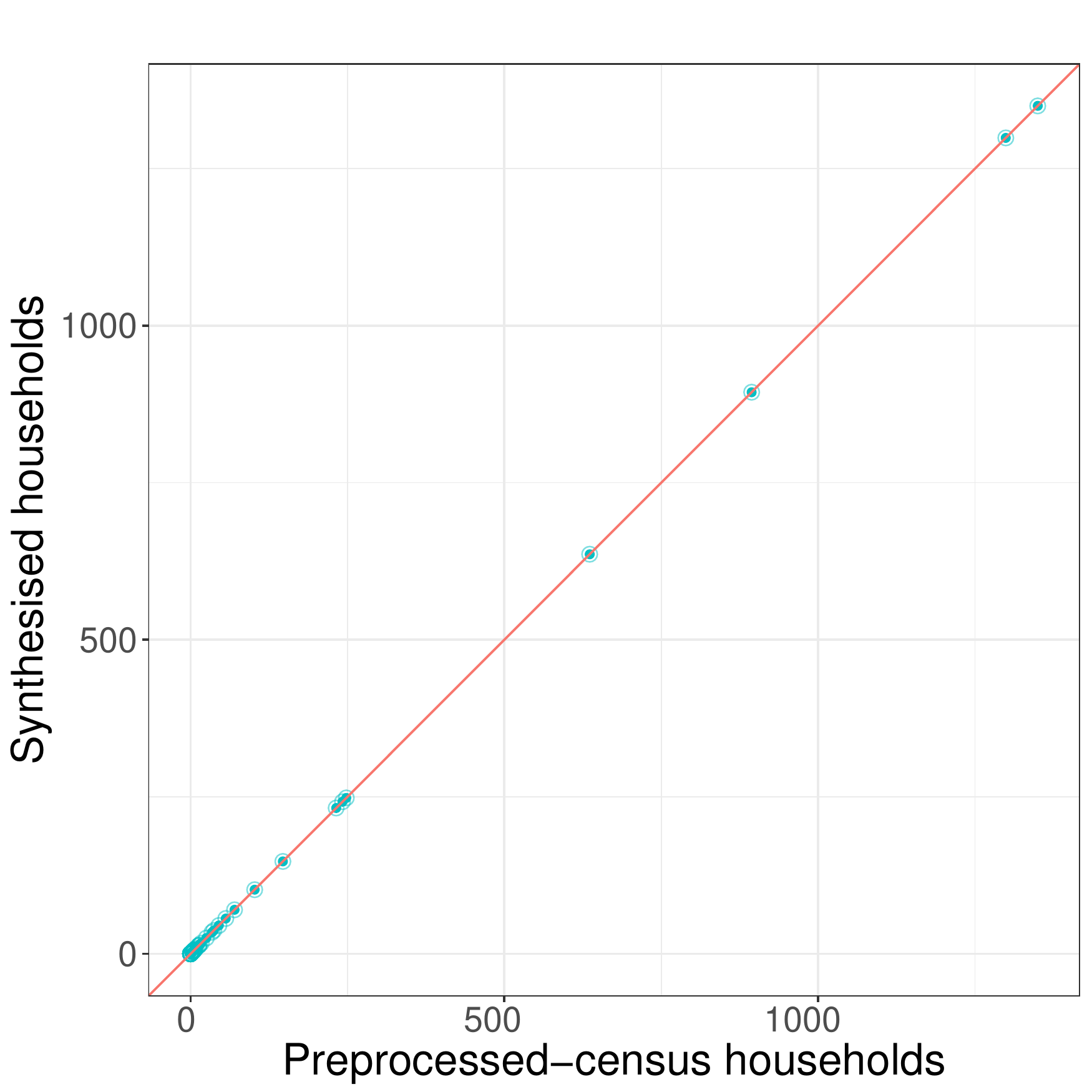}
  }
  \hfill
  \subfloat[Age\label{subfig:yarraville_qq_age}]{
    \includegraphics[width=0.3\textwidth]{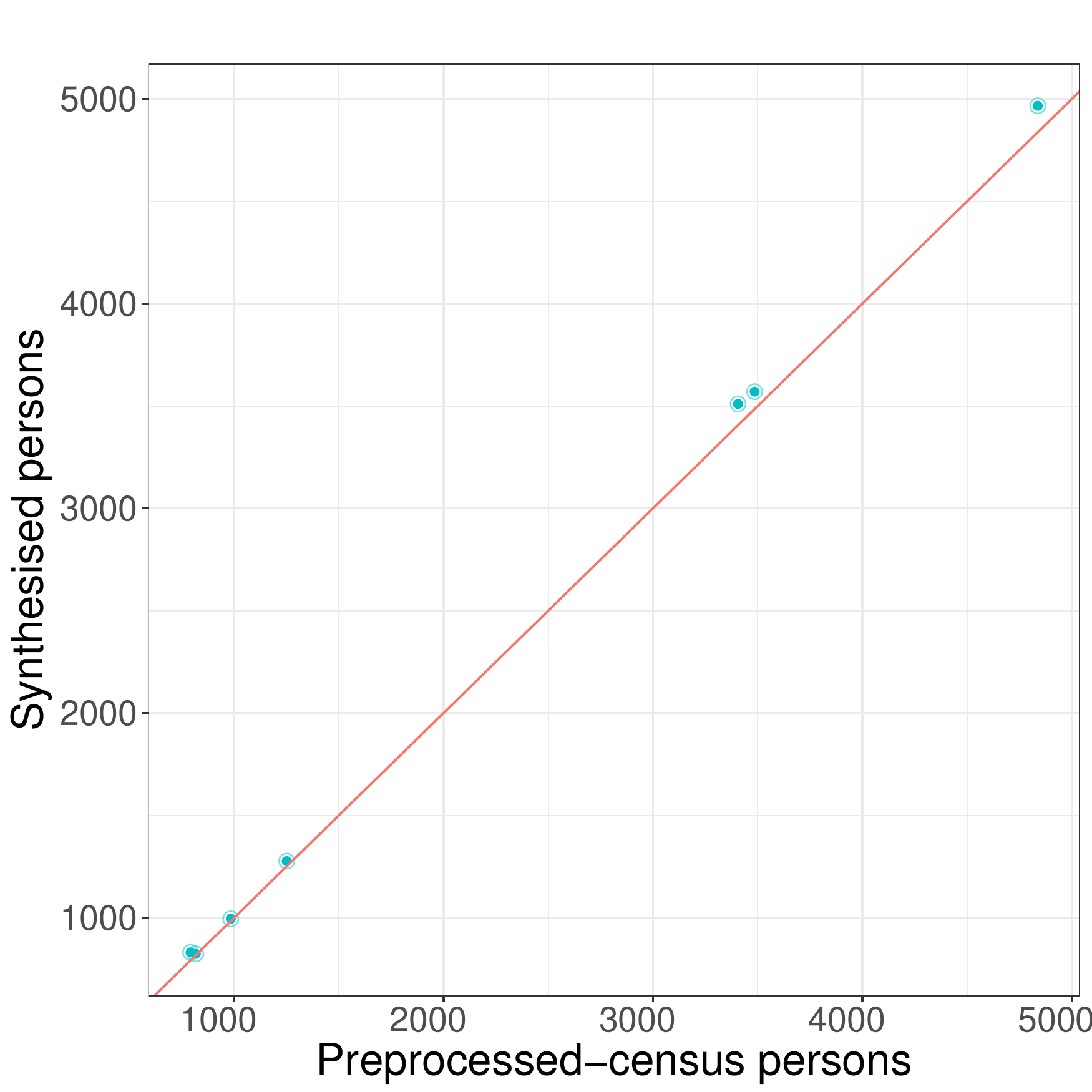}
  }
  \caption{Similarity of Yarraville SA2 synthesized population to
    cleaned census data with respect to persons-level,
    household-level, and age categories; blue dots indicate numbers of
    the synthesized population for each category, with the diagonal
    line showing what would be a perfect match.}
  \label{fig:yarraville-qq}
\end{figure}

A visualisation of the similarity achieved in an example SA2 of
Yarraville that has a synthetic population of 13926 persons in 5533
households in the 2016 census, is given in
figure \ref{fig:yarraville-qq}. The diagonal line in these figures represents
a perfect match, while the blue dots plot actual numbers for each
category in the synthesized population. Figure
\ref{subfig:yarraville_qq_person} shows the 90 person categories,
figure \ref{subfig:yarraville_qq_HH} shows the 65 household
categories, while figure \ref{subfig:yarraville_qq_age} shows the 7
age categories\footnote{Some dots are on top of each other and can't
  therefore be individually distinguished.}. As can be seen, all
points are close to the diagonal,  
indicating strong similarity, as also indicated by the cosine
similarity scores of 1 for households, 0.99 for person categories and
0.97 for persons by age,  for this SA2.

\Omit{
Figure~\ref{fig:sample-sa2-comparison} illustrates the size of the
error in the synthetic population in terms of numbers, for the 
Yarraville SA2. Each blue dot
indicates the census count (y-axis) for a given category (x-axis)
giving 90 person-level points (Figure~\ref{subfig:sa2-person-error}),
65 household-level points (Figure~\ref{subfig:sa2-hh-error}) and 104
age points (Figure~\ref{subfig:sa2-age-error}). For each census point
we plot a corresponding red circle whose radius gives the absolute
count difference (error) for that category in the synthetic
population.

\begin{figure}[ht]
  \subfloat[Person-level error\label{subfig:sa2-person-error}]{
    \includegraphics[width=0.3\textwidth]{analysis/zero-person-types-removed/2016/R/Yarraville_qqplot_persons_preprocessed_vs_synthetic.pdf}
  }
  \hfill
  \subfloat[Household-level error\label{subfig:sa2-hh-error}]{
    \includegraphics[width=0.3\textwidth]{analysis/zero-person-types-removed/2016/R/Yarraville_qqplot_households_preprocessed_vs_synthetic.pdf}
  }
  \hfill
  \subfloat[Age error\label{subfig:sa2-age-error}]{
    \includegraphics[width=0.3\textwidth]{analysis/zero-person-types-removed/2016/R/Yarraville_qqplot_age_cats_persons_preprocessed_vs_synthetic}
  }
  \caption{Yarraville SA2 population with respect to persons-level, household-level, and age categories; blue dots indicate census counts per category; corresponding red circles give extent of error for the category in the synthetic population.  }
  \label{fig:sample-sa2-comparison}
\end{figure}
}
Three SA2s that performed poorly (cosine $<$0.9) were Braeside, Melbourne
Airport, and Port Melbourne Industrial that had very small resident
populations of less than 1\% of the average SA2 population of 10,000
(Table~\ref{tab:latch:cosine-weak-sa2s}), and had more pronounced
inconsistencies between person and
household distributions.
Performance is poor for these SA2s because with low counts there are limited
options in each pool making it difficult to find a good match.

%
%
%
%

\section{Comparison to other techniques} \label{sec:comparison}

To further evaluate our approach, we compare it to two techniques: the
well-known sample-based Iterative Proportional Updating (IPU)
based synthesis 
method, and the sample-free approach of~\cite{Huynh2016} to which our work is
more closely aligned.

\subsection{Evaluation against IPU based approach}

We created the synthetic population for Melbourne using Iterative Proportional
Updating (IPU) based method proposed in \cite{Ye2009}
that synthesises a
population with both person-level and household-level attributes like our
algorithm in Section~\ref{subsec:latch-algo}, but instead uses a disaggregated
sample. Our intent here was to understand how our sample-free algorithm
compares to a state-of-the-art sample-based technique. For this exercise, we
used the IPU implementation published in the Urban Data Science
Toolkit\footnote{\url{https://github.com/UDST/synthpop}}.

Since a disaggregated sample was unavailable for the 2016 census, we use the
previous 2011 Australian census data for which this data could be accessed. The
sample available to us represented
1\% of each SA4 area in Greater Melbourne, where a SA4 typically contains over 100,000 persons as per the ABS hierarchy. To generate the population using IPU for each SA2 for Melbourne, we used the SA4 data corresponding to the SA2s we wanted to include.

 The 2011 census data differed to the 2016 data described in
 Section~\ref{subsec:census-ind-data}~and~\ref{subsec:census-hh-data}. For 2011 at
 person level we only used 84 categories, being 6 relationship status
 categories (\texttt{\married, \loneparent, \children, \relative,
 \grouphh, \loneperson}) $\times$ 2 gender types $\times$ 7 age categories with
 \texttt{85 or over} as the oldest age category. For household level we only
 had 60 categories, being 6 size categories (\texttt{[1,2,\ldots,6]})
 $\times$ 10 household
unit types (\loneperson $+$ \grouphh $+$ [1/2-\family $\times$ 4 primary
family types]).

We ran IPU as per~\cite{Ye2009} with 84-category person-level and
60-category household-level input distributions for 278 SA2s in
Melbourne\footnote{The 2011 census data lists 278 SA2s for Melbourne,
  as compared to 309 in 2016.} for 20,000 IPU iterations to allow it
to converge to 0.0001 quality of fit. For each SA2, the program
generated 20 populations instances and output the one with highest
goodness of fit. We then ran the 
Cosine similarity test on the IPU generated population and our population.

\Omit{
\begin{table}
  \centering%
  \caption{Comparative results for the 2011 Melbourne population of 278 SA2s}%
  \label{tab:latch:ipu-results}
  \begin{tabular}{|p{0.25\textwidth}|p{0.20\textwidth}|p{0.20\textwidth}|p{0.15\textwidth}|p{0.15\textwidth}|}
    \hline
    \multirow{2}{*}{\bf Distribution} &
    \multicolumn{2}{l|}{\bf SA2s with FT $p$-value $<$ 0.05} &
    \multicolumn{2}{l|}{\bf SA2s with CS $>$ 0.99}\\
    ~ &
    \multicolumn{1}{l}{\bf IPU} & \multicolumn{1}{l|}{\bf Ours} &
    \multicolumn{1}{l}{\bf IPU} & \multicolumn{1}{l|}{\bf Ours}
    \\\hline\hline
        Person-level & 14 (5.0\%) & {\bf 8} (2.9\%) & 274 (98.6\%) & {\bf 274} (98.6\%)\\
        Household-level & 156 (56.1\%) & {\bf 0} (0\%) & 277 (99.6\%) &
        {\bf 278} (100\%)\\
        Age & 36 (12.9\%) & {\bf 35} (12.6\%) & 275 (98.9\%) &
       {\bf 275} (98.9\%)\\
        \hline
    \end{tabular}
\end{table}
} 

\begin{table}
  \centering%
  \caption{Comparative results for the 2011 Melbourne population of 278 SA2s}%
  \label{tab:latch:ipu-results}
  \begin{tabular}{|p{0.25\textwidth}|p{0.20\textwidth}|p{0.20\textwidth}|}
    \hline
    \multirow{2}{*}{\bf Distribution} &
    \multicolumn{2}{l|}{\bf SA2s with CS $>$ 0.99}\\
    ~ &
    \multicolumn{1}{l}{\bf IPU} & \multicolumn{1}{l|}{\bf Ours}
    \\\hline\hline
        Person-level &  274 (98.6\%) & {\bf 274} (98.6\%)\\
        Household-level  & 277 (99.6\%) & {\bf 278} (100\%)\\
        Age &  275 (98.9\%) & {\bf 275} (98.9\%)\\
        \hline
    \end{tabular}
\end{table}

Table~\ref{tab:latch:ipu-results} shows the quality of output of our
algorithm compared to IPU on the 2011 census data. 
Both algorithms produce a very good match ($>$98.6\%)
on all three kinds of distributions tested. 
We note that run times for both algorithm were comparable with IPU taking 107s and our algorithm 150s to complete.

\subsection{Comparison to Huynh et. al~\cite{Huynh2016}}

Our work is influenced by~\cite{Huynh2016} that also uses a sample-free approach, uses a population heuristic as we do, and also applies to Australia, albeit to the Sydney region. Importantly, we note that in starting out we had discussions with the authors (prior to their work being published), were given access to their algorithm code, and had intended to use their approach adapted for Melbourne.
We clarify here the reasons why we developed an alternative approach. First, the algorithm in~\cite{Huynh2016} was designed for 2006 census data, whereas we wanted to work with the more recent 2011 data and the upcoming 2016 data that included some structural changes. Second, at that stage their existing work~\cite{NamaziRad2014a,Huynh2013} used IPFP to construct the input joint distributions, which was no longer necessary given these could now be directly obtained using ABS TableBuilder. Third, their code at that stage was closed, and an open implementation that we could contribute to was some way away. We therefore decided to continue developing the work we had started, in parallel to work reported in~\cite{Huynh2016}. We completed the work in 2016, and have spent significant effort since in making it usable Australia-wide.

Our approach differs from~\cite{Huynh2016} as follows:
\begin{enumerate*}[label=(\alph*)]
\item we use new SA2 areas introduced in the 2011 census, whereas they use CCD areas which were discontinued after 2006;
\item we generate multi-family households as per census data, while they assume households to have only one family and add members to it to fit household size;
\item we use exact age distributions
from census, whereas they assign ages uniformly within each age bin which is less accurate;
\item we determine age gaps between couples using a configurable heuristic while they use a Gaussian distribution (in both cases actual data is unknown).
\end{enumerate*}

A direct comparison to~\cite{Huynh2016} is not possible due to different census years used. For 2006 data~\cite{Huynh2016} show that
the distribution of number of households\footnote{Household and family are interchangeable in~\cite{Huynh2016}.} by family composition types matches perfectly for all CCDs. For the number of
households by size, about 10\% of CCDs are different.
For  person-level, the number of males/females by age and relationship status are different for 10\% of CCDs.
Our algorithm, for 2011 data, matches households by number and size perfectly, while person-level differences are 
less than 2\%
(Table~\ref{tab:latch:ipu-results}).

\section{Conclusion} \label{sec:conclusion}

We presented a sample-free approach to population synthesis in the Australian context, that uses heuristics for building the population of individuals and assigning them to households in dwellings. The algorithm is fast in that it can generate the full population of Melbourne with 4.5 million persons in 1.8 million households in less than three minutes on a recent-day computer. The algorithm produces a perfect match for all 65 household-level
distributions for the 2016 and 2011 census data. For person-level categories, we find greater than 98\% similarity for 2016 and 2011 census data when using the Cosine Similarity (CS) test.

%
%
The reason why our algorithm matches households perfectly is because
our heuristic treats household distributions as the reference and
adjusts person-level attributes to fit. The reason we did this was
that our applications were focussed on households rather than
individuals. 
It would be straightforward to adjust this bias to spread it between
the two data sets, or to regard the person level data as the
reference, by slightly modifying the
cleaning process which ensures consistency between the two data
sets. This decision should ne made based on the application for which
the population is being constructed.

%

\bibliographystyle{splncs04}
\bibliography{arXiv.bib}

\end{document}